\documentclass[sigconf,nonacm]{acmart}

\setcopyright{acmlicensed}
\copyrightyear{2026}
\acmYear{2026}
\acmDOI{XXXXXXX.XXXXXXX}
\acmConference[CCS '26]{the 2025 ACM SIGSAC Conference on Computer and Communications Security}{November 15--19,
  2018}{Netherlands}
\acmISBN{978-1-4503-XXXX-X/2018/06}
\acmSubmissionID{123-A56-BU3}

\AtBeginDocument{%
  \providecommand\BibTeX{{%
    \normalfont B\kern-0.5em{\scshape i\kern-0.25em b}\kern-0.8em\TeX}}}

\usepackage{amsmath, amsfonts, amsthm, amssymb}
\usepackage{algorithmic}
\usepackage{graphicx}
\usepackage{textcomp}
\usepackage{xspace}
\usepackage{enumitem}
\usepackage{xurl}
\usepackage{balance}

\usepackage{adjustbox,multirow}

\usepackage{tcolorbox}
\newtcolorbox{lightgraybox}{
  colback=gray!10,%
  colframe=gray!40,%
  boxrule=0.5pt,%
  arc=2mm,%
  left=3pt, right=3pt, top=3pt, bottom=3pt,%
}

\newtcolorbox{darkgraybox}{
  colback=gray!25,%
  colframe=black,%
  boxrule=0.6pt,%
  arc=2mm,%
  left=4mm, right=4mm, top=2mm, bottom=2mm,%
}

\usepackage[labelfont=bf,skip=5pt,figurename=Fig.]{caption}

\usepackage{subcaption}

\usepackage{listings}
\usepackage{xcolor}
\usepackage[T1]{fontenc}

\definecolor{light-gray}{gray}{0.95}%
\definecolor{comment-gray}{rgb}{0.41,0.6,0.33}%
\lstdefinestyle{mystyle}{
    language=Python, 
    captionpos=b,%
    basicstyle=\fontfamily{phv}\selectfont\footnotesize,%
    keywordstyle=\bfseries,%
    breaklines=true, 
    breakatwhitespace=true, 
    showstringspaces=false, 
    numbers=left, 
    numberstyle=\tiny\color{black}, 
    numbersep=10pt, 
    commentstyle=\color{comment-gray},%
    backgroundcolor=\color{light-gray},%
    xleftmargin=17pt, 
    xrightmargin=3.4pt, 
    framextopmargin=15pt,%
    framexbottommargin=10pt,%
    framexleftmargin=17pt,%
    framesep=3pt, 
    fillcolor=\color{light-gray},%
    linewidth=\linewidth, 
    basewidth=0.5em,
    deletekeywords={[2]file},
    morekeywords={match, if, then, label_host, label_file, drop, alert, allow, declassify, endorse, contains},%
    aboveskip=2ex,
    belowskip=2ex,
    escapechar=|
}

\def\BibTeX{{\rm B\kern-.05em{\sc i\kern-.025em b}\kern-.08em
    T\kern-.1667em\lower.7ex\hbox{E}\kern-.125emX}}

\usepackage{amsmath}
\usepackage{tikz}
\newcommand*\circled[1]{\tikz[baseline=(char.base)]{
            \node[shape=circle,fill,inner sep=1pt] (char) {\textcolor{white}{#1}};}}

\usepackage{booktabs}
\usepackage{float}
\usepackage{makecell}

\usepackage{graphicx}
\definecolor{strongred}{RGB}{190, 0, 0}

\usepackage{subfiles}
\usepackage{hyperref}

\usepackage[capitalize,nameinlink]{cleveref}
\hypersetup{%
	bookmarksnumbered, bookmarksopen=true, bookmarksopenlevel=1,%
}
\crefname{figure}{Fig.}{Figs.}
\crefname{listing}{Listing}{Listings}
\crefname{section}{Section}{Sections}
\crefname{table}{Table}{Tables}
\crefname{BNF}{Grammar}{Grammars}
\crefname{algorithm}{Algorithm}{Algorithms}
\crefname{line}{Line}{Lines}

\AtBeginEnvironment{appendices}{\crefalias{section}{appendix}\crefalias{subsection}{appendix}}

\usepackage{soul}

\newcommand{\cmtt}{\fontfamily{cmtt}\selectfont}
\newcommand{\cmss}{\fontfamily{cmss}\selectfont}

\newcommand{\tool}{\textsc{NetZone}\xspace}

\newcommand{\object}{{\cmss AccessScope}\xspace}
\newcommand{\objects}{{\cmss AccessScopes}\xspace}

\renewcommand{\paragraph}[1]{\smallskip\noindent\textbf{#1.}}

\newcommand{\PP}[1]{\vspace{2px}\noindent{\underline{\textbf{#1}}}}

\begin{document}

\title{Breaking Ambient Trust: In-Network Per-Process Access Control Against Lateral Movement}

\author{Osama Bajaber}
\affiliation{%
  \institution{King Abdulaziz University}
  \city{Jeddah}
  \country{Saudi Arabia}
}
\email{obajaber@kau.edu.sa}

\author{Bo Ji}
\affiliation{%
  \institution{Virginia Tech}
  \city{Blacksburg}
  \state{VA}
  \country{USA}
}
\email{boji@vt.edu}

\author{Peng Gao}
\affiliation{%
  \institution{Virginia Tech}
  \city{Blacksburg}
  \state{VA}
  \country{USA}
}
\email{penggao@vt.edu}

\begin{abstract}
Enterprise networks remain vulnerable to Advanced Persistent Threats (APTs), where adversaries gain an initial foothold and move laterally across the network, accumulating access permissions hop by hop to reach critical targets.
Existing network defenses cannot track user movement at the process level across the network; instead, they \emph{grant ambient trust} to all processes within a host. 
As a result, once a host is compromised, malicious processes inherit the victim’s permissions, \emph{thereby expanding the attacker’s access scope} and enabling further lateral movement.

To address this gap, we present \tool, an in-network access control that confines each user process to a fixed access scope that persists as the user moves across the network. 
\tool introduces a new abstraction, called \object, which represents a lightweight access capability bound to the user’s processes. 
Each \object encodes the set of hosts a user identity is authorized to access and is embedded in the process’s outgoing network traffic for validation before reaching its destination. 
As users pivot across hosts, \object propagates with their traffic, rebinds to the receiving process, and persists across hosts.
This ensures that \emph{regardless of network location}, the user’s processes are consistently governed by their bound \object and their access permissions remain unchanged. 
To handle the high volume of network traffic generated by processes, we develop a data-plane co-design that integrates programmable switches with eBPF. 
\tool employs a set of in-network optimizations and lightweight \object persistence techniques to inspect the embedded \object on the fly, enabling line-rate processing of high traffic volumes with negligible latency overhead. 
Our extensive evaluations show that \tool can effectively defend against various sophisticated attack scenarios without introducing noticeable overhead.

\end{abstract}



\maketitle

\section{Introduction}
\label{sec:intro}

Modern enterprise networks remain vulnerable to Advanced Persistent Threats (APTs)~\cite{apt}.
These attacks share a common structure: an adversary gains an initial foothold and then moves laterally across hosts, progressively accumulating access permissions beyond their entry point (e.g., a compromised host becomes a stepping stone to internal and privileged systems).
It is through this incremental accumulation of permissions that APTs achieve their goals, leading to major data breaches~\cite{data_breach}, significant financial losses~\cite{financial_loss_1}, and widespread service disruptions~\cite{service_disruption_1}.
Despite decades of investments in network access control~\cite{network_access_trend}, attackers still succeed as existing network defenses rely on coarse-grained, network-level information (i.e., IP addresses and port numbers) and lack visibility to track user process movement across hosts. 
As a result, they cannot enforce consistent permissions as users move.

\begin{figure*}[t]
     \centering
     \begin{subfigure}[b]{0.45\textwidth}
         \centering
         \includegraphics[width=\textwidth]{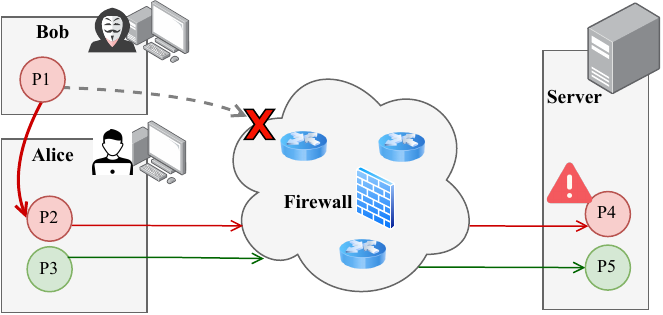}
         \caption{Stepping-stone attacks leverage intermediate hosts to hide the attacker's origin and reach a final target.}
         \label{fig:motivating_example_attack}
     \end{subfigure}
     \hfill
     \begin{subfigure}[b]{0.45\textwidth}
         \centering
         \includegraphics[width=\textwidth]{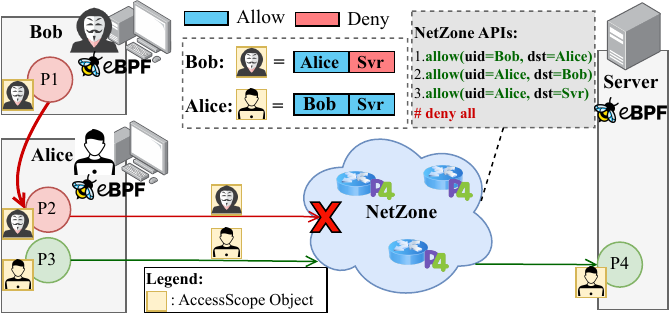}
         \caption{\tool enforces user-specific access scopes via process-level \object objects.}
         \label{fig:motivating_example_netzone}
     \end{subfigure}
        \caption{An enterprise network where a firewall blocks Bob from reaching the Server but allows Alice (an administrator) to access it. In (a), Bob cannot directly connect to the Server (dashed gray arrow) but exploits a vulnerability on Alice’s system to spawn a malicious process {\cmtt P2}, using it as a stepping stone to bypass the firewall. In (b), \tool encodes each user’s permissions in \object objects (yellow boxes) that are bound to processes and \emph{persist} across hosts. \tool parses \objects within programmable switches and enforces permissions at line rate, regardless of the connection’s origin or the user’s location.}
\vspace{0ex} 
\label{fig:motivating_example}
\end{figure*}

Consider a representative scenario in~\cref{fig:motivating_example_attack} with two hosts (Alice and Bob) and a sensitive server.
A firewall permits only Alice (an administrator) to access the server.
Although Alice and Bob can communicate with each other, Bob cannot directly reach the server.
However, Bob turns malicious and uses Alice as a stepping stone to reach the server by sending a spear-phishing email that installs a backdoor on her machine.
Leveraging this backdoor, Bob establishes a remote session using process {\cmtt P1} and spawns a new process {\cmtt P2} on Alice’s host.
Since the firewall enforces only coarse IP- and port-based rules, it cannot distinguish between traffic from Bob’s malicious process {\cmtt P2} and Alice’s process {\cmtt P3}, allowing Bob to inherit Alice's permissions and pivot toward the server.
Similar attacks arise through various other techniques (e.g., access token theft~\cite{mitre_access_token}, session hijacking~\cite{mitre_session_hijacking}, or exploitation of shared services~\cite{mitre_shared_services}), emphasizing a broader and pervasive security challenge.

This exposes two fundamental weaknesses in current network defenses: (1) the \emph{ambient trust implicitly granted to all processes within a host}, and (2) the inability to \emph{track user identity at the process-level} as users move across the network.
Defenses like distributed firewalls
enforce \emph{coarse, IP- and port-based rules} rather than the identity of the process 
initiating the connection.
As a result, when an attacker pivots onto a compromised host, all outgoing connections appear to originate from the legitimate user, and malicious processes inherit the same access permissions.
Even recent identity-aware solutions~\cite{ethane, netview} that incorporate user roles ultimately translate high-level policies into coarse-grained IP-based rules for enforcement, thereby trusting all processes on a host and failing to track user movement at the process level across hosts.
This limitation is critical in data centers and cloud environments, where thousands of processes interact through shared infrastructure, enabling attackers to expand permissions and pivot deeper into the network~\cite{data_breach, microsoftbreach, snowflakebreach}.

\paragraph{Goal \& challenges}
What is needed is a fundamentally different model where access permissions are tied to the user’s identity rather than the host’s network location, and follow the user as they move across the network.
Our key insight is twofold: (1) \textbf{each process should carry distinct access permissions} based on the user’s identity; (2) these permissions should \textbf{persist and propagate with the user as they move across hosts}, ensuring that a user maintains consistent access permissions regardless of network location.

To realize this insight, \textbf{our idea} is to bind a \emph{lightweight credential object} to each user’s processes.
This credential specifies which hosts a process is authorized to access and is \emph{attached} to outgoing network packets for verification at every access attempt.
This credential also \emph{propagates} as the user pivots across hosts and spawns new processes.
Through this mechanism, regardless of where the user lands, access decisions are governed by the credential object, strictly limiting the user's access scope to authorized destinations.

As illustrated in~\cref{fig:motivating_example_netzone}, credential objects encode administrator-defined policies that specify each user’s authorized access.
Even if Bob successfully compromises Alice’s system, his malicious process {\cmtt P2} carries a restricted credential object that \emph{does not} permit access to the server.
In contrast, Alice’s legitimate process {\cmtt P3} holds a different credential object that authorizes the connection.

However, realizing this vision introduces three \textbf{key challenges}.
\begin{enumerate}[leftmargin=*, itemsep=1pt, topsep=1pt]

    \item \emph{Compact policy representation.} Designing a credential object that encodes a user’s permissions across all hosts is non-trivial.
    The object must remain compact to propagate efficiently while still expressing permissions for potentially thousands of hosts.
    
    \item \emph{Cross-host persistent.} Permissions must remain bound to each process and propagate as users move across hosts, bridging intra-host activities (i.e., spawning new processes and threads) with inter-host communications (i.e., network connections), ensuring credentials propagate both within and between hosts.
    
    \item \emph{Line-rate enforcement at scale.} Validating credentials on every access attempt in a high-volume enterprise network must operate in real time without introducing noticeable overhead or degrading benign traffic.

\end{enumerate}

\paragraph{Contributions}
We introduce \tool, an in-network access control system that enforces per-user, per-process access scopes across the enterprise network.
\tool builds on 
capability-based access control~\cite{capstone} and brings this notion of fine-grained, decentralized access control to enterprise networks.
Unlike existing capability-based systems~\cite{sel4, cheri} that are restricted to a single host, \tool extends capability enforcement across the network to regulate process access to remote hosts.
At the core of \tool is a new abstraction called the \object object.
This object encodes the set of hosts a user identity is authorized to access and is bound to the process identifier (PID).
A process must be presented whenever the process attempts to access a remote host.
As users spawn new processes or pivot across hosts, their associated \object \emph{propagates} with them, preserving a consistent access scope throughout the network.
This is achieved by embedding the \object associated with the sending process’s PID into outgoing packets, and propagating the corresponding capability to the recipient process.

The key advantage of this design is that binding the \object to individual processes eliminates the ambient trust granted to all connections from the same host.
Instead, \tool preserves each user’s access scope as they move across the network, enforcing consistent access permissions regardless of network location.
Even if a host is compromised, malicious processes cannot inherit access permissions of legitimate ones and pivot.
This decentralized, process-level design \emph{confines} each user to a specific set of authorized hosts, effectively mitigating stepping-stone attacks.

To efficiently enforce \objects carried on network packets and propagate them across hosts, \tool employs a novel \emph{data plane co-design} of programmable switches and eBPF.
Programmable switches (e.g., Tofino) offer data plane programmability and guarantee customized Tbps line-rate packet processing.
This enables \tool to create custom packet headers that carry \object objects and inspect them directly in the data plane at line rate.
We further developed optimization techniques that enable efficient \object generation, distribution, validation, and enforcement across the network.
To bind \objects to processes and propagate them within and between hosts, we developed lightweight kernel programs using eBPF technology.
eBPF enables the operating system (OS) kernel to be extended through sandboxed custom programs,
without modifying kernel source code or user-level applications.
Our lightweight eBPF components efficiently trace process activity, bind processes to their corresponding \objects within hosts, and attach or extract \objects from network packets to ensure seamless propagation.

\paragraph{Evaluations}
We extensively evaluated \tool on both a physical testbed with a Tofino programmable switch and on realistic large-scale topologies using packet-level simulations, incorporating both synthetic and real-world workloads.
Our results show that:
(1) \tool can prevent a wide range of advanced attacks that evade existing network defenses~\cite{iptable, ethane, netview}, while adding only a negligible 140~ns latency to enforce \object on the fly.
(2) \tool scales to real-world traces, including DARPA OpTC~\cite{darpa}, Yatesbury~\cite{netvigil}, and LANL Unified Host and Network~\cite{lanl} datasets, achieving 99.9~Gbps throughput on a 100~Gbps (per-port) programmable switch with minimal impact on legitimate traffic.
(3) The eBPF programs add only 3-9~$\mu s$ to the total time of the monitored kernel events with a small 130~MB memory footprint, imposing negligible performance impact on hosts.
(4) To further demonstrate the benefits of our data-plane design, we compare \tool with centralized identity-based network access control solutions~\cite{ethane, netview}. 
\tool enforces policies 27-60x faster while remaining resilient against saturation attacks.

\section{Background}
\label{sec:motivation_background}

\paragraph{Network-access control}
Existing network access control solutions can be broadly grouped by the granularity of identity they enforce. However, each group leaves critical gaps that enable attackers to gain additional permissions.

Traditional firewalls~\cite{iptable} are widely deployed and offer fast prevention with low compute overhead.
However, they enforce low-level IP- and port-based rules, with no visibility into the identity of the process behind a connection.
Access scopes are therefore tied to network location: any process running on a permitted host implicitly inherits that host's access permissions. 
An attacker who lands on a permitted host can freely abuse those inherited permissions to move laterally without triggering a firewall rule.

Signature-based firewalls~\cite{pigasus, snort_offloader, kargus} attempt to enrich rules with deep packet inspection (DPI). 
Yet, much of today’s application-layer traffic is encrypted (e.g., HTTPS, TLS), limiting their ability to extract meaningful user identity content.
More fundamentally, DPI still treats each host as a single trust unit, leaving ambient trust unaddressed.
DPI is also computationally expensive, often requiring dedicated CPU~\cite{pigasus, snort_offloader} or GPU~\cite{kargus} resources, imposing significant performance overhead on the network.

A more recent line of work~\cite{ethane, fml, netview} 
extends software-defined network (SDN) architectures with role- and attribute-based access control.
These systems use high-level policies to determine network reachability based on the user’s identity behind each host.
While this brings identity into the policy definition, enforcement still ultimately reduces to IP-based 
rules installed in the data plane.
As a result, trust is granted to the \emph{host} rather than to \emph{individual processes}, and permissions are enforced at the host level rather than re-evaluated for each process initiating a connection.
This means that once a host is compromised, the attacker's processes appear indistinguishable from the legitimate user's.
Furthermore, these systems mediate every connection through a centralized controller, introducing significant per-connection latency and a single point of failure vulnerable to saturation attacks.

To date, no solution fundamentally eliminates ambient trust by enforcing per-user, per-process access scopes that persist as users move laterally across the network.

\paragraph{Capability-based systems}
Capabilities~\cite{capstone} provide an efficient and secure mechanism for fine-grained resource management and protection.
They enable decentralized access control by granting entities a token that defines a set of access permissions.
A process can access a resource only by presenting the corresponding capability.
This security model limits the resources permitted to each holder and enforces the principle of least privilege~\cite{saltzer1975}, ensuring that the holder can access only the resources explicitly authorized.

Many recent works have introduced capability-based operating systems~\cite{sel4, cheri}, which regulate what resources a process can access. 
However, these systems operate strictly within OS boundaries.
Their design focuses primarily on regulating access to local resources (e.g., files, memory locations, and I/O devices) rather than extending protection to accessing network hosts.
They also require extensive kernel modifications to implement capability-based access control, which significantly limits their adoption.
In contrast, \tool extends the capability-based model beyond OS boundaries to regulate process-level access to remote hosts, protecting against stepping-stone attacks.
Moreover, \tool does not require any modifications to the OS kernel or running applications to enforce capabilities.
This allows seamless deployment and broad applicability of our defense system.

Capabilities have been used for many security purposes, such as in programming languages~\cite{joe2010, prog_cap} and providing access confinement in capability-based operating systems~\cite{sel4, cheri}.
However, these solutions are restricted to a single host.
Other efforts have integrated capabilities to regulate access to distributed resources (e.g., IoT~\cite{iot_cap1, iot_cap2}, and cloud computing~\cite{cloud_cap1,cloud_cap2}).
Unlike \tool, these systems operate at a coarse granularity and assume ambient trust for all processes on the same host.

\paragraph{Programmable data planes}\label{sec:programmable_data_plane}
Programmable switches have emerged as high-performance, Tbps-rate dataplane targets enabled by languages like P4~\cite{bosshart2014}.
A P4 program defines the packet headers and protocols to be supported, along with a programmable parser that extracts these headers from incoming packets.
The extracted headers are represented as a packet header vector (PHV) and processed through a pipeline of hardware stages, each performing match/action operations.
Within each stage, match/action tables use keys (e.g., header fields) to look up entries and execute corresponding actions such as arithmetic, bitwise, or metadata updates.
Arithmetic Logic Units (ALUs) in these stages perform simple computations, while the programmable deparser later reconstructs the packet before forwarding it to the next hop.
However, these memory resources are limited, typically on the order of O(100)MB for SRAM and O(10)MB for TCAM.
These constraints fundamentally restrict what can be processed at line rate, posing a significant challenge in designing efficient and scalable data-plane solutions.

\section{System Overview}
\label{sec:overview}

\begin{figure}[t]
    \centering
    \includegraphics[width=\linewidth]{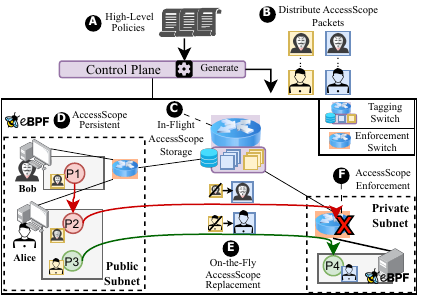}
    \caption{\tool system workflow.
    }
    \label{fig:system_overview}
\end{figure}

One key challenge is keeping \objects lightweight to be efficiently carried with user traffic and processed within programmable switches.
A naive approach would be to generate a single \object per user.
However, a single \object may need to encode permissions for thousands of hosts, making it too large.
Instead, \tool partitions \object into smaller, \textbf{per-subnet} \objects, with each representing the user’s permissions within a specific network subnet.
This design aligns naturally with enterprise network structure, where subnets (e.g., VLANs and ToR domains) partition large networks into smaller domains for easier management~\cite{subnet}.
A user carries only the \object corresponding to their current subnet.
When the user moves to a new subnet, \tool automatically replaces the current \object with the corresponding one for the destination subnet.

Even with compact \objects, efficient in-network enforcement remains challenging.
Programmable switches, while capable of line-rate packet processing, have limited on-chip resources (i.e., SRAM, TCAM, and processing capacity).
Maintaining per-user \objects and enforcing per-packet access decisions are both resource-intensive operations that together cannot fit within a single switch.
This necessitates a division of labor, which \tool adopts by separating these functions into two switch roles.
\textbf{Tagging switches} are gateway switches positioned at the boundary of each subnet. 
These switches attach the appropriate per-subnet \objects to each user’s connection before the traffic enters the subnet.
\textbf{Enforcement switches} run within subnets, focusing solely on validating \objects and enforcing security actions at line rate.

\cref{fig:system_overview} provides an overview of \tool.
We now briefly describe the high-level workflow of the defense.

\textbf{Initialization: } This phase outlines how administrators define high-level policies and how the control plane generates and distributes the corresponding \objects to tagging switches when a user logs in.

\begin{enumerate}[label={\textbf{(Step \arabic*)}}, leftmargin=*,itemsep=1pt, topsep=1pt]

\item \tool provides a set of APIs (see~\cref{sec:operational_deployment}) that allows administrators to define high-level, identity-centric policies through the control plane \protect\circled{A}.
As shown in~\cref{fig:motivating_example_attack}, these policies specify which hosts a particular user is allowed to access.

\item When a user logs in, our eBPF programs on the user's system retrieve the UID from the running OS. It then sends the UID to the \tool control plane to trigger the generation of the user’s \object.

\item The control plane uses the received UID to retrieve the user's high-level policies and encodes them into a per-subnet \object that defines the user’s access permissions (more details at~\cref{subsubsec:generating_object}).

\item The control plane then distributes the generated \objects using the lightweight \object protocol (see~\cref{sec:netscope_protocol}) to store them in the tagging switches \protect\circled{B}. 
\tool also sends the user’s current subnet \object to their host system to bind existing processes with the received \object.

\item Each tagging switch stores the \objects that define access permissions for hosts in its local subnets (see~\cref{sec:in-flight_caching}) \protect\circled{C}.

\end{enumerate}

\textbf{Propagating \objects: } This phase describes how \tool persists \objects across hosts.

\begin{enumerate}[label={\textbf{(Step \arabic*)}}, leftmargin=*,itemsep=1pt, topsep=1pt]
\setcounter{enumi}{5}

\item When a process initiates a connection to another host, the eBPF program \protect\circled{D} on the sender constructs a custom \object packet header (see~\cref{fig:packet_header}) that carries the user's UID and access permissions. 
The eBPF program then attaches the process's \object to the outgoing traffic.

\item Upon receiving the packet, the eBPF program on the receiving host extracts the appended \object and binds it to the receiving process.

\item The eBPF programs within the destination host maintain the received \object by binding it to any new processes and threads spawned by the user, preserving the same access permissions.

\item If the user initiates a connection to a different subnet, the tagging switch at the boundary intercepts the connection \protect\circled{E}.
The tagging switch extracts the appended UID in the connection, looks up the corresponding destination-subnet \object from its local \object store, and replaces the user’s current \object with the destination-subnet \object before forwarding the connection into the target subnet.

\end{enumerate}

\textbf{Enforcing \object: } This phase describes how \tool enforces \objects entirely within the switch data plane.

\begin{enumerate}[label={\textbf{(Step \arabic*)}}, leftmargin=*,itemsep=1pt, topsep=1pt]
\setcounter{enumi}{9}

\item Before a connection reaches its destination within a subnet, the enforcement switch intercepts the packet carrying the \object.

\item The switch inspects the \object \protect\circled{F} to verify whether it authorizes access to the destination host and then forwards or drops the packet accordingly.

\end{enumerate}

In~\cref{sec:operational_deployment}, we further describe how \tool reactively updates \objects when policies change, handles packet loss, and recovers from switch failures.

\paragraph{\tool in action}
\cref{fig:system_overview} illustrates how \tool prevents the attacker in our motivating example.
Network administrators define high-level policies via \tool APIs to restrict access to the private subnet to administrators (i.e., Alice), while regular users (i.e., Bob) are confined to the public subnet.
Both Alice’s and Bob’s processes are initialized with their respective \objects.
When Bob’s malicious process {\cmtt P1} compromises Alice’s system and spawns a new malicious process {\cmtt P2}, Bob’s \object propagates along this connection to {\cmtt P2}.
Later, when both the malicious process {\cmtt P2} and the benign process {\cmtt P3} attempt to connect to the server, the tagging switch replaces their public-subnet \objects with new ones specifying each user’s permissions in the private subnet.
The enforcement switch then inspects the \object for each flow: it blocks {\cmtt P2}, which carries Bob’s restricted \object, and allows {\cmtt P3}, which carries an authorized \object.

\paragraph{Threat model}\label{sec:threat_model}
We consider attackers who gain an initial foothold into the network either externally from the public Internet or as insiders within the network.
After compromising a host, the attacker can spawn processes or threads, initiate and receive connections, and laterally move across hosts by exploiting legitimate network paths. 
We consider that the programmable switches and control plane are secure as they are under the control of the network administrators.
Also, we assume that our eBPF programs cannot be tampered with, which can be further secured by recent tamper-proof and tamper-evident auditing techniques~\cite{custos}. 
Adversaries with kernel-level access could disable our eBPF programs and any other running security mechanisms, which is beyond the scope of this work and falls under kernel security rather than network security.
We assume network administrators define appropriate high-level policies to regulate per-user access permission.
We focus on enforcing these policies consistently as users move across hosts, rather than correcting the policies themselves.
Our threat model aligns with established previous works on network security~\cite{ethane, netview} and defenses leveraging programmable switches~\cite{qiao2020, Jiarong2020netwarden, p4control, jaqen2021, poseidon2020, mew2023}.

In~\cref{sec:operational_deployment}, we further demonstrate how \tool is resilient against adaptive adversaries attempting to compromise the system itself.

\section{Decentralized \object Architecture}
\label{sec:design}

In this section, we detail the \object abstraction, the tagging switch primitive, and the techniques we develop to replace \object across network subnets in real time.

\subsection{\object Abstraction}\label{sec:netscope_abstraction}
\object is an object that encodes a user's permitted access in a network subnet.
A naive design would define a single \object for each user to encode their entire set of permissions across the network.
However, such monolithic objects are not scalable: they can grow excessively large and become impractical for network switches with limited resources to process them.
Instead, we generate a distinct \object \emph{for each network subnet}.
Each \object is uniquely defined by the tuple:
\begin{equation} \label{eq:object_definition}
\centering
\object =(UID,Subnet\_ID,Permissions)   
\end{equation}

\begin{itemize}[leftmargin=*,itemsep=1pt, topsep=1pt]

\item \PP{UID}: This represents the user identity as defined in the identity and access management (IAM) system managed by network administrators (e.g., Active Directory).
Each UID is unique within its IAM domain and remains consistent for a user across all hosts managed under that domain.
In multi-domain environments (e.g., multi-forest Active Directory or federated IAM setups), \tool adds the IAM domain identifier to the UID to guarantee global uniqueness across domains.

\item \PP{Subnet\_ID}: To uniquely identify network subnets, \tool derives $\mathrm{Subnet\_ID}$s from existing network isolation mechanisms, such as VLANs and subnets in traditional environments, and Virtual Private Clouds (VPC) in cloud environments.
Specifically, \tool's control plane retrieves the VLAN and routing information from network switches, extracts the defined network subnets, and assigns a unique $\mathrm{Subnet\_ID}$ for each subnet in the network.
Each $\mathrm{Subnet\_ID}$ defines the subnet where a given \object specifies the user’s allowed access scope.
The control plane maintains a canonical mapping of network subnet IP-to-$\mathrm{Subnet\_ID}$, ensuring each IP address is consistently associated with the correct subnet.

\item \PP{Permissions}: Encodes which hosts within a $\mathrm{Subnet\_ID}$ the user may access.

\end{itemize}

\tool dynamically updates and revokes $\mathrm{Subnet\_ID}$s and permissions as hosts migrate or reconfigure, ensuring that \object objects remain consistent with the current network state (more details in~\cref{sec:operational_deployment}).

The user’s complete policy across the network is the union of all their \object objects:

\begin{equation} \label{eq:object_union}
\centering
N_u = \left\{ (Subnet\_ID_i, Permissions_i) \mid UID = u \right\}
\end{equation}

The set $N_u$ constitutes the user’s permissions across all subnets. 
When a user’s connection is about to enter subnet $i$, \tool tags the user's connection with \object that matches its UID and the $\mathrm{Subnet\_ID}$ the user is entering as \object $\big(UID, Subnet\_ID_i, Permissions_i\big)$.
Within the subnet, if the destination host is not present in $Permissions_i$, the connection is blocked.

\subsection{\object Distribution}\label{sec:in-network_caching}

\tool leverages programmable switches to enforce user-specific access scopes via \objects.
Suppose we make the user carry their union set of \objects (representing all resources in the network). 
This incurs high transmission overhead and forces the user to carry many permissions \emph{irrelevant to the user’s current network subnet.}
Alternatively, preloading all switches with every user’s \objects is infeasible due to limited on-chip memory, especially in large enterprise networks with thousands of users.

To address this, \tool adopts an \textbf{\emph{in-network division of labor}}.
Gateway switches at the boundary of each network subnet act as \emph{tagging switches}, while switches inside the subnet act as \emph{enforcement switches}.
When a user gets authenticated, the control plane generates the user's \object objects for all network subnets.
The control plane then distributes them via our lightweight protocol (details in~\cref{sec:netscope_protocol}) in the form of packets.
Each tagging switch stores only the \objects that are relevant to resources in its directly connected subnet.
As the user laterally moves from one network subnet to another, the tagging switch at the destination subnet \emph{replaces} the source-subnet \object with the destination-subnet \object.
Inside each subnet, enforcement switches (detailed in~\cref{sec:in-network_enforcement}) inspect the attached \object and mediate access to local resources.

Next, we present the design of our custom \object packet header, explain how \tool generates user-specific \objects from high-level policies, and describe how the \object distribution protocol delivers \objects to tagging switches.

\begin{figure}[H]
    \centering

    \begin{subfigure}[b]{0.45\textwidth}
        \centering
        \includegraphics[width=\linewidth]{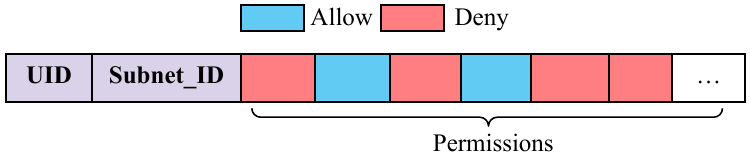}
        \caption{\object structure}
        \label{fig:object_struct}
    \end{subfigure}
    
    \vspace*{1.2em}%

    \begin{subfigure}[b]{0.45\textwidth}
        \centering
        \includegraphics[width=\linewidth]{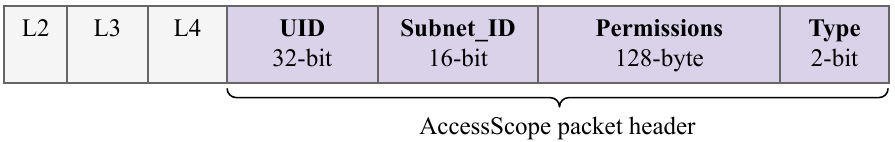}
        \caption{\object packet header}
        \label{fig:packet_header}
    \end{subfigure}  

    \caption{An illustration showing \object structure and our custom packet header}
    \label{fig:packet_header_object_struct}
\end{figure}

\subsubsection{\object Custom Packet Header}

\cref{fig:object_struct} illustrates \object structure.
The UID and $\mathrm{Subnet\_ID}$ specify the sender’s identity and the subnet where the \object applies.
The permission field is a bit string, where each bit represents access to a host in the subnet.
If the bit is set to one, the user is allowed to access the corresponding host; otherwise, the user is blocked.
The permission field size depends on the subnet size.
In large-scale clusters such as Google’s~\cite{alfares}, a single ToR switch typically serves hosts within a /22-/24 IP subnet (corresponding to 254-1022 hosts), making a 128B (1024-bit) permission field sufficient to represent all hosts in a network subnet.
We detail how \tool maps each bit position to its corresponding host in~\cref{subsubsec:generating_object}.

To carry \object in the network, \tool employs a \emph{custom packet header} (see~\cref{fig:packet_header}).
We set the reserved bit in the IP fragmentation field~\cite{evilbit} to mark packets carrying our \object packet header. 
Our custom packet header can appear in two forms:
\begin{enumerate}[leftmargin=*, itemsep=1pt, topsep=1pt]
    \item \PP{Standalone \object control packet:} Generated by the control plane (not part of the user's data traffic) to carry \objects across the network and store them in tagging switches.
    
    \item \PP{Embedded \object packet headers}: Added to user data packets to carry the user’s \object to be inspected by enforcement switches.
\end{enumerate}

The \textbf{Type} field is used to distinguish these roles:
\begin{itemize}[leftmargin=*,itemsep=1pt, topsep=1pt]
    \item \PP{(Type=0):} Newly generated standalone \object control packets by the control plane for distributing and storing.
    \item \PP{(Type=1):} Standalone \object control packets stored in the tagging switch and ready to be assigned as the user enters a new network subnet.
    \item \PP{(Type=2):} User data packets carrying an embedded \object packet header, which are validated by enforcement switches.
\end{itemize}

A naive way of carrying embedded \object packet headers is to attach them to every user data packet. 
This unnecessarily increases packet size, since the same security decision applies to all packets within a user’s connection.
Instead, \tool attaches the \object header only to the \emph{initial packets} of a connection.
This is sufficient for enforcement switches (see~\cref{sec:object_enforcemnt}) to inspect the \object packet header once and apply the same decision to all subsequent packets (which do not carry a \object packet header) in the connection.
Also, \tool incorporates robust mechanisms to handle missing \object packets through ACK packets and restores state after switch failures (see~\cref{sec:operational_deployment}).

\subsubsection{Generating \objects from High-Level Policies}\label{subsubsec:generating_object}
\tool control plane translates high-level policies (defined via our APIs in~\cref{sec:operational_deployment}) into \objects.
When a user logs in, our eBPF program on the host sends a packet containing the user’s UID to trigger the control plane to generate the corresponding \objects.
The control plane then fetches the high-level policies for the received UID and extracts the permitted host IPs.

To generate \objects, the control plane takes the user’s UID, permitted hosts, and the IP-to-$\mathrm{Subnet\_ID}$ mapping (see~\cref{sec:netscope_abstraction}).
For each $\mathrm{Subnet\_ID}$, it encodes permitted hosts as a bit in the permission bit string. 
The result is a set of per-subnet \objects, each containing the UID, $\mathrm{Subnet\_ID}$, and its permission bit string (as shown in~\cref{fig:object_struct}).
Furthermore, the control plane creates a dictionary for each $\mathrm{Subnet\_ID}$ that maps permission bit positions to host IPs.
These dictionaries are loaded into the enforcement switches’ match/action tables to interpret permission bits for destination IPs.
When a user's policy is updated, \tool automatically updates affected \objects (details in~\cref{sec:operational_deployment}).

\subsubsection{\object Distribution Protocol}\label{sec:netscope_protocol}
After generating \objects, the control plane distributes them to tagging switches so they are available when the user enters a subnet.
For each \object, \tool creates a standalone control packet (Type=0) that carries the header to a specific $\mathrm{Subnet\_ID}$.
To distribute these packets efficiently, \tool employs the \textbf{\emph{spanning tree protocol}} among tagging switches.
The protocol builds a loop-free topology across switches, preventing \object control packets from circulating indefinitely.
The control plane elects a root switch, and each switch configures its forwarding ports to forward \object packets along the tree.
When a tagging switch receives a \object control packet, it checks whether the carried $\mathrm{Subnet\_ID}$ matches a locally connected network subnet. 
If so, the switch marks the packet as (Type=1) and stores it. 
Otherwise, the packet is forwarded to downstream switches in the tree.
This approach ensures that all \object control packets reach their corresponding tagging switch without flooding the network.

\subsection{In-Flight \object Storage}\label{sec:in-flight_caching}

As users enter network subnets, the tagging switch fetches the user’s \object and attaches it to their connection, ensuring it is enforced within the destination subnet (see~\cref{sec:in-network_enforcement}).
A key challenge, however, is that tagging switches must keep \objects readily available to tag incoming user traffic.
In large-scale environments such as Google clusters~\cite{google_clusters1, google_clusters2}, this can involve up to 10K active users.
Storing \objects for this many users in switch memory is infeasible due to two main constraints:
\begin{enumerate}[leftmargin=*, itemsep=1pt, topsep=1pt]
    \item Limited on-chip memory (hundreds of MBs of SRAM and tens of TCAM).
    
    \item Fixed-width entries (up to 64 bits~\cite{tofino}), which cannot accommodate entire \objects (up to 128 bytes) or their variable-length permission fields.
\end{enumerate}

A recent line of work on in-network caching~\cite{in-network_caching} introduced a caching architecture that overcomes the limitations of constrained switch memory in traditional in-network key-value stores~\cite{netcache, farreach}.
This architecture supports storing variable-length cache items by keeping them as in-flight packets within the data plane.
This design leverages the built-in feature of programmable switches that allows packets to re-enter the recirculation pipeline through an internal loopback port (separate from other data packet ports).
This design has proven effective in expanding switch storage capacity and handling variable-length cache items for load-balancing applications without increasing the switch’s memory footprint.

Building on this idea, \tool adopts a similar architecture to keep \objects in-flight within the switch data plane.
When a standalone \object control packet (Type=0) arrives at a tagging switch that must store it, the switch modifies its type field to indicate stored state (Type=1) and sends the packet to the internal loopback port.
The switch then matches incoming user connections (Type=2) against these in-flight \objects, selecting the correct one based on the user’s UID as they enter the subnet.
\tool employs failure-handling mechanisms that coordinate with the control plane if an \object is missing or a switch fails (see~\cref{sec:operational_deployment}).

\begin{figure}[t]
    \centering
    \includegraphics[width=\linewidth]{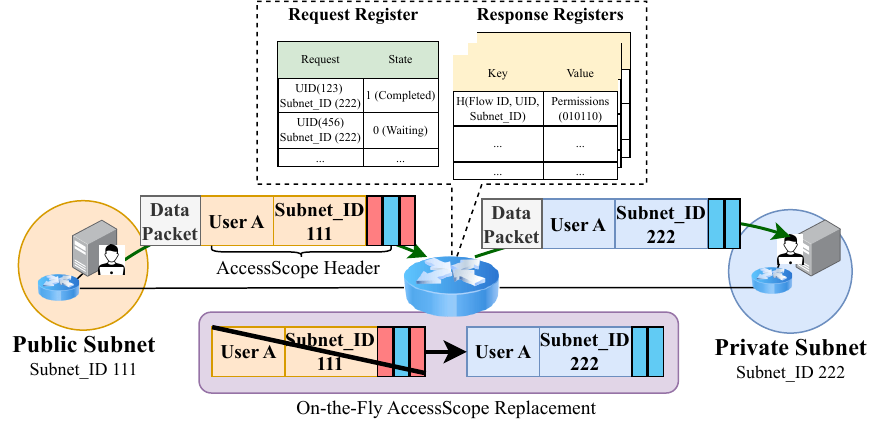}
    \caption{Replacing \object as the user moves to a new network subnet}
    \label{fig:header_replacement}
\end{figure}

\subsection{On-the-Fly \object Replacement}\label{sec:on-the-fly_tagging}

When a user’s connection enters a new network subnet, the tagging switch needs to retrieve the user's \object.
However, since \objects are stored in-flight as packets (Type=1), programmable switches cannot directly copy an \object header from the stored in-flight packet and attach it to the user’s connection.
To resolve this, \tool implements a \emph{lightweight request-response coordination mechanism} within the data plane using two stateful registers: {\cmtt requestReg} and {\cmtt responseReg}, as illustrated in~\cref{fig:header_replacement}.

\paragraph{Request register}
When an initial user packet carrying an \object header (Type=2) reaches a new subnet, the switch determines the destination $\mathrm{Subnet\_ID}$ from the packet’s destination IP. 
Consequently, the switch adds an entry in {\cmtt requestReg}, keyed by {\cmtt (UID, $\mathrm{Subnet\_ID}$)}.
This records that there is a request to get the \object for the user's UID for $\mathrm{Subnet\_ID}$.
The user's data packet is recirculated until the request is resolved.
Meanwhile, in-flight \object control packets (Type=1) check {\cmtt requestReg} for matching entries.
If a match is found, the switch responds to the request and writes the matched stored \object values to {\cmtt responseReg}, indexed by the connection’s 5-tuple ($\mathrm{IP_{src}}$, $\mathrm{Port_{src}}$, $\mathrm{IP_{dst}}$, $\mathrm{Port_{dst}}$, $\mathrm{Protocol}$), UID, and the destination-$\mathrm{Subnet\_ID}$.

\paragraph{Response registers}
A challenge in storing \object permissions in {\cmtt responseReg} is their variable sizes.
Each \object encodes a bitmap that can grow up to 1024 bits, while switch registers store only fixed-width words (32-64 bits).
To address this, \tool splits each permission field into fixed-size chunks and stores them sequentially across multiple {\cmtt responseReg} registers, indexed by the flow’s 5-tuple, UID, and $\mathrm{Subnet\_ID}$.
The waiting user’s data packet then replaces its existing \object header with a new one for the destination subnet, reconstructing the complete \object by sequentially reading and combining the matching entries from these registers.
In practice, since the \object is attached only to the initial packets of a connection, this process adds less than 100$\mu$s of delay to a single waiting packet, which is negligible given that the RTT in typical enterprise networks spans several milliseconds. 
Subsequent packets are forwarded without modification with a negligible $\sim$130ns delay (see~\cref{sec:eval_system_impact}).

\section{Enforcing and Persisting \objects}\label{sec:in-network_enforcement}

Once the user’s \object packet header enters a network subnet, \tool enforces the encoded access scope at line rate and persists this scope as the user moves within the subnet.

\subsection{In-Network \object Enforcement}\label{sec:object_enforcemnt}

Enforcement switches mediate all connections within a subnet.
\tool leverages the embedded \object packet header (Type=2) in the user’s initial packet to restrict access to authorized hosts.
Each bit in the permission field represents access to a specific host.
To interpret these bits efficiently, \tool employs \emph{bitwise operations} in the data plane.
The switch maintains a match/action table (see~\cref{sec:in-network_caching}) that maps each host IP address to a bitmask, where the bit representing the destination host is set to 1 and all others to 0.
When a user's process initiates a connection to a host, the switch retrieves the destination bitmask and performs a bitwise {\cmtt AND} with the \object permission field.
A nonzero result indicates the process is authorized; otherwise, the connection is dropped.

\paragraph{Cache security decision} 
Since the \object is carried only in the initial packet of a connection, \tool enforces the same security decision to subsequent packets in the connection that do not carry the header.
After \tool inspects the packet carrying the \object (Type=2), the decision is stored in {\cmtt DecisionTable}, indexed by the connection 5-tuple key and the decision value (0 to block, 1 to allow). 
Subsequent packets match against this table and are processed accordingly.
To support various protocols, \tool adopts the following strategies: the \object header is attached to the SYN packet in TCP handshakes, to the first few packets in new UDP flows until an \object ACK is received, and to both request and reply packets in ICMP exchanges.

\subsection{Cross-Host \object Persistence}\label{sec:cross-host_propagation}

Within a subnet, a user may move across hosts and perform various operations such as spawning new processes and threads.
Enforcing least privilege requires tracking intra-host activity at the process level rather than treating each host as a single trust boundary.
This requires a lightweight mechanism to track and propagate identity context as processes and threads are created, without intrusive kernel changes or application-level modifications.

\tool achieves this using eBPF programs that propagate \objects within hosts and across the network.
eBPF~\cite{ebpf} is a widely adopted kernel technology~\cite{ebpf_enterprise} that provides a secure sandbox for running user-defined programs inside the operating system kernel.
These programs attach to kernel hooks (e.g., system calls or network events), enabling custom functionality without changing the kernel source or loading new modules.

\textbf{\textit{(1) Extracting \object from incoming packets:}}
Once a user reaches a host over the network, we design an eBPF program that extracts the \object packet header from incoming packets.
For high-performance processing, this program runs at the XDP (eXpress Data Path) hook~\cite{xdp2018}, enabling execution directly in the network interface card driver.
Packets marked with the reserved IP fragment bit are parsed to extract and remove the \object header before the packet is passed to the kernel stack.
The program then stores the extracted \object and the destination port in a BPF map called {\cmtt incomingObject (dport->\object)}.

To bind the incoming \object to the receiving process or thread, we attach a eBPF program to trace system calls used to accept the connection (i.e., {\cmtt accept()} and {\cmtt accept4()} for TCP, and {\cmtt recvfrom()} and {\cmtt recvmsg()} for UDP).
Once such a system call is detected, our eBPF program looks up the destination port in {\cmtt incomingObject} and, if matched, retrieves the process or thread identifier using {\cmtt bpf\_get\_current\_pid\_tgid()}.
This helper returns a 64-bit value that combines the process ID (upper 32 bits) and thread group ID (lower 32 bits), enabling fine-grained tracking even when threads spawn new processes.
The extracted \object is then stored in a BPF map {\cmtt objectMap (pid\_tgid->\object)}.

\textbf{\textit{(2) In-kernel \object propagation:}}
To track how a user’s identity propagates within a system, we attach multiple eBPF programs to system calls that create new execution contexts.
Specifically, when a process spawns a child via {\cmtt fork()}, {\cmtt clone()}, or {\cmtt execve()}, the eBPF program captures the parent’s {\cmtt pid\_tgid}.
The parent's {\cmtt pid\_tgid} is then used as a lookup key in {\cmtt objectMap} to retrieve its associated \object.
Consequently, the eBPF program creates an entry in {\cmtt objectMap}, indexed by the child’s {\cmtt pid\_tgid}, ensuring that the child inherits the correct \object.

\textbf{\textit{(3) Propagate \object back to the network:}}
When a user initiates a connection to another host in the network, our eBPF program intercepts the corresponding system calls (i.e., {\cmtt connect()} for TCP and {\cmtt sendto()} for UDP).
The program extracts the caller's {\cmtt pid\_tgid} and records the mapping in a BPF map {\cmtt outgoingObject (sport->\object)}.
To attach \object packet header to outgoing packets, we attach an egress eBPF program at the Traffic Control (TC) hook.
It checks whether the outgoing packet's source port is in {\cmtt outgoingObject}.
If found, the eBPF program constructs an \object packet header and inserts it into the matched outgoing packet.

\section{Handling Practical Requirements}\label{sec:operational_deployment}

\paragraph{Adaptive adversaries}
\tool is resilient to adaptive adversaries targeting our system itself.
An attacker on the victim's host may attempt PID spoofing to gain another process's \object.
However, this attack fails since the OS enforces unique PIDs for all active processes.
A more advanced man-in-the-middle attacker may try to capture \objects in transit and manipulate their permissions to gain unauthorized access.
To prevent this, \tool uses in-network keyed hashing~\cite{hashing2019} and integrates a lightweight message authentication code (MAC) scheme.
Specifically, our eBPF programs compute a keyed hash over each \object packet header and append it to the outgoing packet, while the switches verify the MAC inline before processing.
Packets with invalid MACs are dropped, preventing forgery or tampering.

An attacker may attempt to saturate the switches by flooding the network with a large number of connections.
However, since \tool processes all packets entirely within the data plane, it remains resilient to such denial-of-service (DoS) attacks (as evaluated in~\cref{sec:eval_comparison}).
Also, an attacker may try to exhaust switch registers by initiating many legitimate connections.
To mitigate this, \tool employs a rate-limiting mechanism that limits the number of connections from each UID within a defined time window.
This allows \tool to restrict malicious processes without disrupting benign processes on the same host. 
Also, \tool periodically evicts inactive entries to free up switch memory.

\paragraph{Failure handling}
\tool employs robust mechanisms to handle packet loss and switch failures.
For packet loss, \tool uses ACKs between tagging switches and the control plane to ensure reliable delivery of \object packets (Type=0).
When a switch stores an \object, it clones the packet and sends it as an ACK to the control plane.
If the control plane does not receive an ACK, the \object is retransmitted until confirmed.

If a tagging switch fails, all in-flight \objects stored on it are lost.
To ensure uninterrupted operations, \tool automatically repopulates stored \objects as follows.
During normal operation, the control plane records which \objects are stored on each tagging switch when ACKs are received.
When a failed switch recovers, the control plane re-sends the corresponding \objects to restore the switch's storage.
This adds negligible storage overhead to the control plane.
In a typical data center cluster with about 10K concurrent active users and 13K hosts~\cite{google_clusters1, google_clusters2}, the control plane needs only $\sim$10~MB of memory to store \objects, which is negligible in modern servers.

\paragraph{Defining high-level policies}
\tool provides a set of APIs for administrators to define user-based policies, which are then encoded into \objects.
These APIs enable easy specification of per-user permissions without requiring low-level data-plane programmability.
An administrator can define a new policy using {\cmtt AddAllowPolicy(UID, dst\_ip)}, granting the specified user access to a destination host.
Similarly, a policy can be removed with {\cmtt RemoveAllowPolicy(UID, dst\_ip)}.
By default, any host not explicitly allowed for a user is denied.

\paragraph{Updating \object}
In dynamic networks with frequent policy changes, \tool supports real-time updates to \object permissions across switches and hosts.
When a policy changes, the control plane generates an updated standalone \object packet (Type=0) and injects it into the network.
The distribution protocol delivers the updated \object to relevant tagging switches while the control plane removes outdated versions.
To synchronize hosts, the control plane sends the same update packets to systems within the updated network subnet, where eBPF programs extract the new \object and overwrite old entries in {\cmtt objectMap}.
This ensures consistent policy enforcement, with updates typically propagating network-wide within 2-10~ms.

\section{Evaluation}\label{sec:eval}

We answer research questions on \tool’s effectiveness, scalability, overhead, and switch resource utilization through extensive evaluations on our physical testbed and using real-world workload datasets and network topologies.

\begin{enumerate}[label={\textbf{(RQ\arabic*)}}, leftmargin=*,itemsep=1pt, topsep=1pt]
  \item How effective is \tool's per-user network scope in mitigating various attacks?
  
  \item How efficiently does \tool process traffic with real-world workloads and network topologies?
  
  \item What impact does \tool have on the network and the host performance?
  
  \item How does \tool’s in-network implementation compare with an alternative server-based implementation?

  \item What is \tool’s resource utilization on programmable switches?
\end{enumerate}

\paragraph{Implementation and experimental setup}
We implement \tool using $\sim$1300 lines of code, including the tagging and enforcement switch primitives, eBPF programs, and switch control plane functions.
For our main experiments, we deployed \tool on a physical Tofino 32x100~Gbps port.
Our setup mirrors existing network security and P4 works~\cite{jaqen2021, poseidon2020, bedrock, smartcookie, Jiarong2021}.
The switch is configured as a Top-of-Rack (ToR) switch, which is connected to three Dell R420 servers, each equipped with an Intel Xeon E5-2430 CPU running at 2.20 GHz, 64~GB RAM, and Ubuntu 22.04.

\begin{figure}[t]
    \centering
    \includegraphics[width=.9\linewidth]{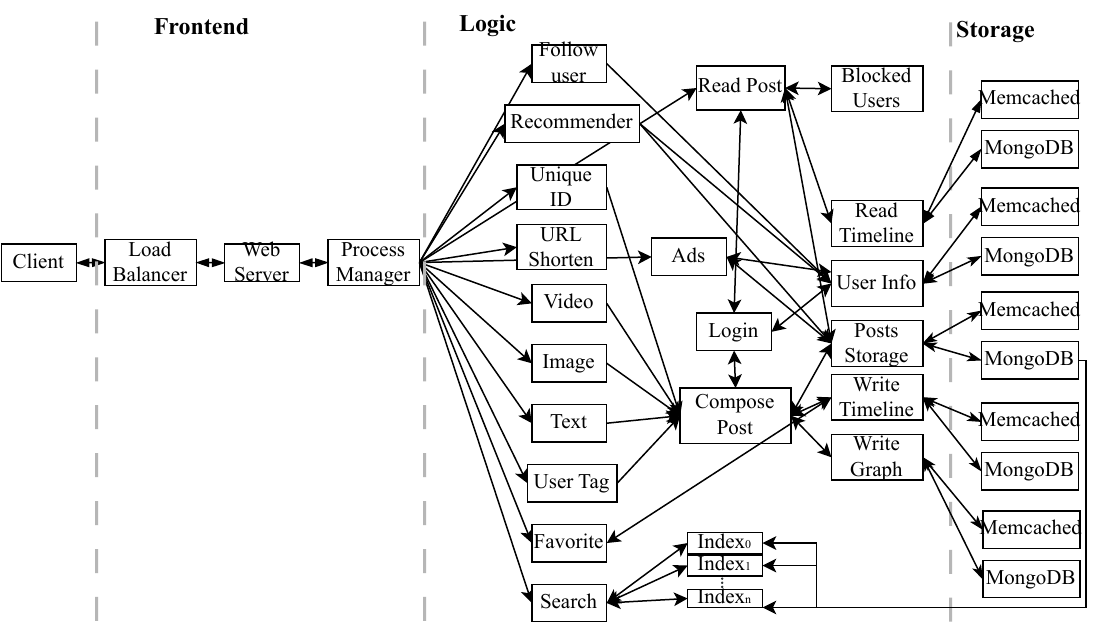}
    \caption{Social media network}
    \label{fig:topology_social_network}
\end{figure}

\begin{figure}[t]
    \centering
    \includegraphics[width=.9\linewidth]{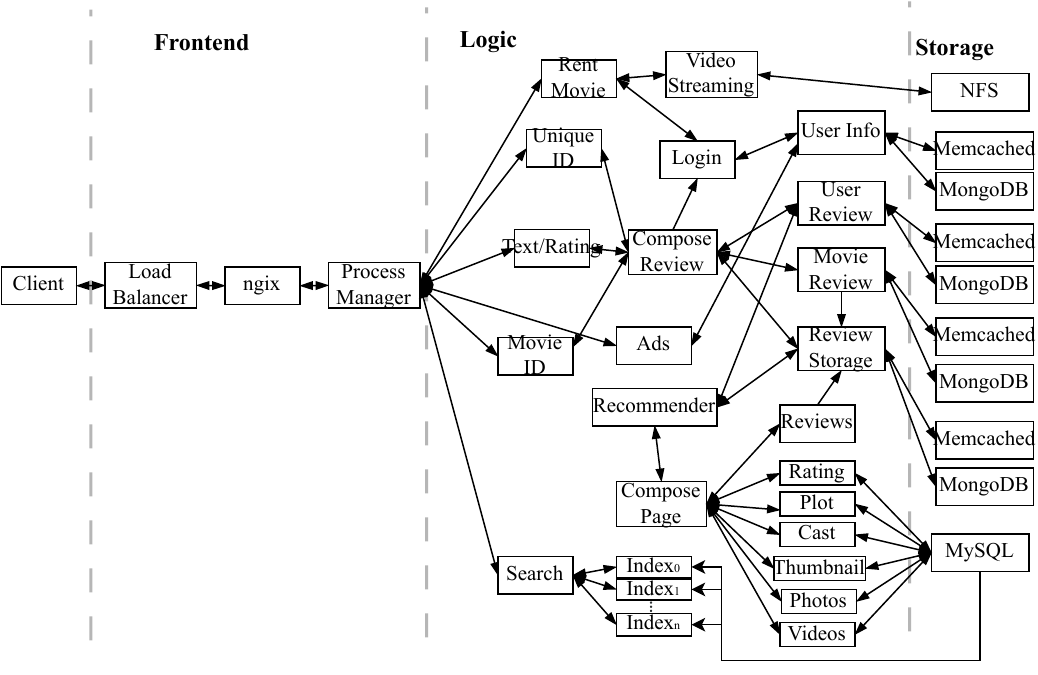}
    \caption{Media streaming service}
    \label{fig:topology_media_service}
\end{figure}

\begin{figure}[t]
    \centering
    \includegraphics[width=.9\linewidth]{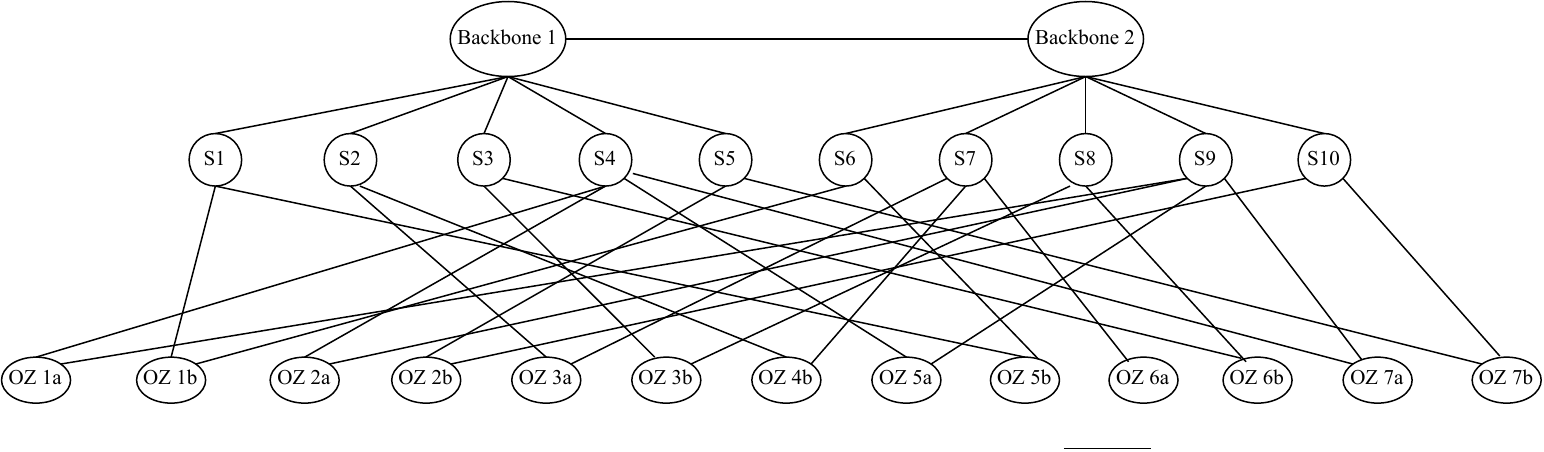}
    \caption{Stanford backbone network}
    \label{fig:topology_stanford}
\end{figure}

We evaluate \tool using three real-world datasets: (1)~DARPA OpTC~\cite{darpa}, (2) Yatesbury~\cite{netvigil}, and (3) LANL Unified Host and Network~\cite{lanl} datasets.
In addition, we deploy \tool in three representative topologies (illustrated in~\cref{fig:topology_social_network,fig:topology_media_service,fig:topology_stanford}) to test against realistic attack scenarios: (1) a web-based social media network~\cite{deathstar}, (2) a media streaming service~\cite{deathstar}, and (3) Stanford University’s backbone network~\cite{stanford_topology}.
The social media network and media streaming services are deployed on a 4-VM scale set across our physical servers, connected through the Tofino switch.
The Stanford backbone network is emulated in Mininet using software P4 switches (bmv2~\cite{bmv2}) to replicate the behavior of physical hardware.

Next, we present three realistic attack scenarios that we conduct on the three topologies.
In each scenario, the attacker employs realistic techniques, such as network scanning, zero-day exploitation, privilege escalation, and lateral movement.

\paragraph{Use case 1: Social media network} 
The architecture (shown in~\cref{fig:topology_social_network}) consists of frontend servers, ad engines, user recommendations, post composition, and metadata storage, with MongoDB and Memcached powering the backend.
We embed a Log4Shell vulnerability~\cite{cve-log4shell} in the frontend web server. 
An attacker compromises the entry point by crafting an HTTP packet header that carries a malicious JNDI lookup string.
This triggers remote code execution, granting the attacker an interactive shell on the compromised frontend.
The attacker then leverages post-compromise tools (e.g., netcat, curl) to pivot laterally toward sensitive backend databases.

\paragraph{Use case 2: Media streaming service}
Its architecture (shown in~\cref{fig:topology_media_service}) includes NGINX frontend servers, a payment authentication module, a streaming service (nginx-hls), and backend databases (MySQL, MongoDB, Memcached).
An attacker targets the movie review field of the web frontend page by exploiting an injection vulnerability~\cite{cve-injection}.
Because the service fails to sanitize user input correctly, the attacker can inject malicious arbitrary code, acquiring an interactive shell.
This enables the attacker to pivot across services, ultimately reaching the backend DB.

\paragraph{Use case 3: Stanford backbone network}
In \cref{fig:topology_stanford}, Stanford University’s backbone network~\cite{stanford_topology} supports 15K students and 2K faculty across 14 operational zones (OZs), interconnected through edge switches and two central backbone switches.
We deployed Infection Monkey malware~\cite{infection_monkey} that initially infects a compromised host.
Then, the malware attempts to propagate laterally across the university’s operational zones by scanning for vulnerable services and exploiting misconfigured or permissive inter-zone connectivity.

\subsection{RQ1: Defense Effectiveness}\label{sec:eval_effectiveness}

We validate the effectiveness of \tool in enforcing per-user access scopes to mitigate realistic attacks.
We deploy our motivating example shown in \cref{fig:motivating_example} in our testbed, implemented on a 3-VM scale to represent each of the running hosts.
We configure a traditional firewall (iptables~\cite{iptable}) that mediates access across two subnets: a Public subnet (hosting Alice and Bob) and a Private subnet (hosting the server).
Firewall rules permit intra-subnet traffic and allow Alice to reach the server.
Direct connections from Bob to the Private Subnet are blocked.
Using the malicious Bob user, we attempt to reach the server by pivoting through Alice as an intermediate host.
We conduct seven attacks, including exploiting a zero-day vulnerability~\cite{distcc} using Metasploit~\cite{metasploit}, deploying Infection Monkey malware~\cite{infection_monkey}, and exfiltrating sensitive data using the Data Exfiltration Toolkit (DET)~\cite{det} over multiple channels (e.g., TCP, UDP, DNS, HTTP, and Gmail).
Each attack is tested with and without \tool, and we measure the attacker’s throughput during the experiments using iperf3.

\begin{figure}[t]
    \centering
    \includegraphics[width=.85\linewidth]{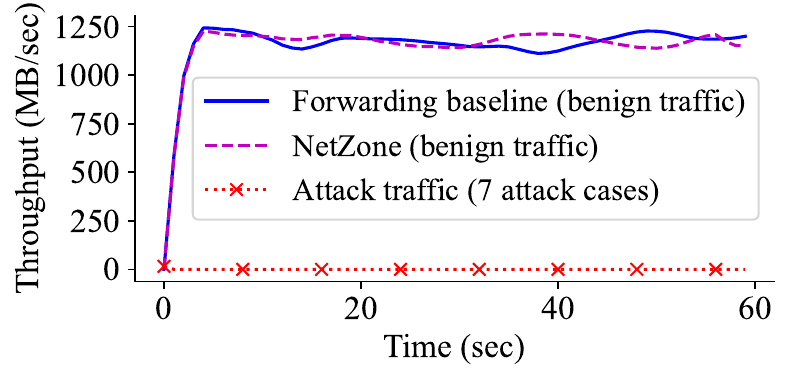}
    \caption{\tool blocks all attack traffic while imposing minimal overhead on benign traffic}
    \label{fig:eval_effectiveness}
\end{figure}

\begin{figure*}[t]
     \centering
     \begin{subfigure}[b]{0.25\textwidth}
         \centering
         \includegraphics[width=\textwidth]{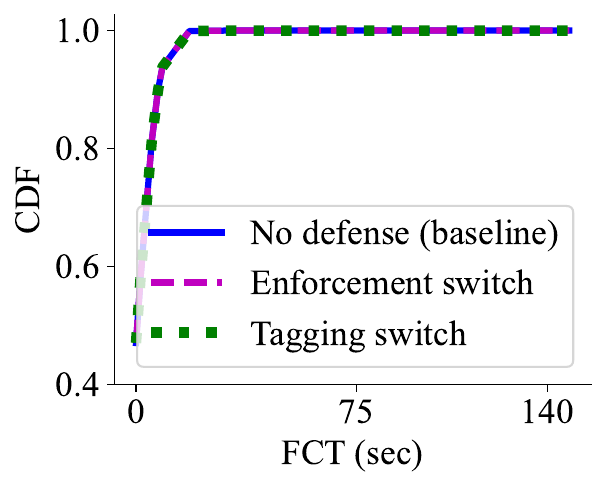}
         \caption{DARPA OpTC FCT}
         \label{fig:darpa}
     \end{subfigure}
     \hspace{1.5em}
     \begin{subfigure}[b]{0.25\textwidth}
         \centering
         \includegraphics[width=\textwidth]{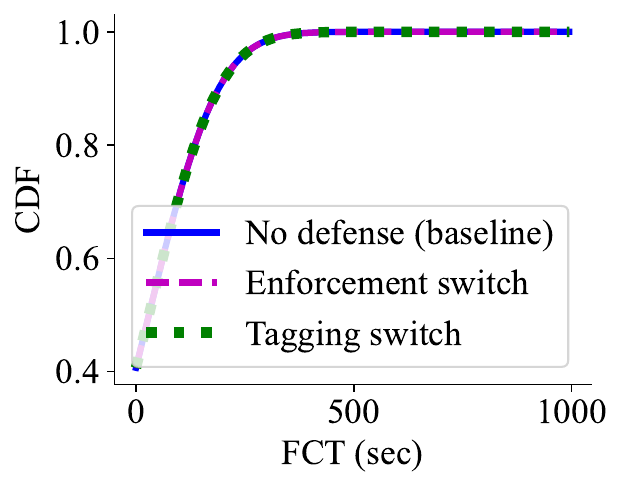}
         \caption{LANL FCT}
         \label{fig:lanl}
     \end{subfigure}
     \hspace{1.5em}
     \begin{subfigure}[b]{0.25\textwidth}
         \centering
         \includegraphics[width=\textwidth]{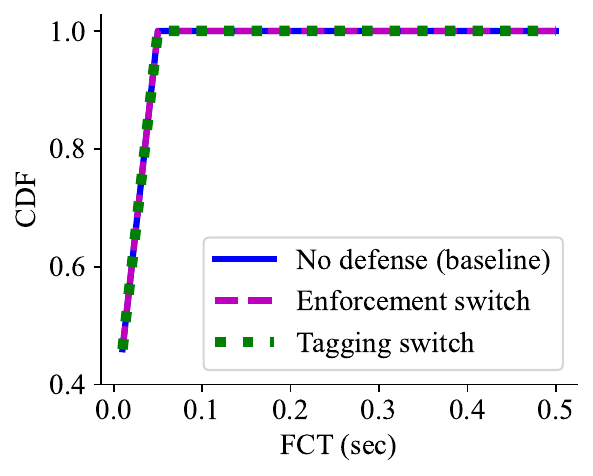}
         \caption{Yatesbury FCT}
         \label{fig:netvigil}
     \end{subfigure}
        \caption{\tool imposes minimal overhead on real-world workloads}
        \label{fig:datasets}
\end{figure*}

\cref{fig:eval_effectiveness} summarizes the results.
Without \tool, the attacker successfully reaches the server and sustains stable throughput for all seven attack variants, enabling data exfiltration despite firewall isolation.
With \tool enabled, we configure high-level policies in the control plane to allow only the user Alice to access the server, while Bob is restricted from accessing the Public Subnet.
The control plane generates per-user \objects accordingly, which are attached to traffic in the network.
In every tested attack, \tool confines the attacker to their permitted access scope and prevents lateral movement. 
Consequently, attacker throughput falls to zero for all seven attacks.

\begin{lightgraybox}
\textbf{Takeaways:} \tool effectively blocks all 7 attacks across 5 communication channels, reducing attacker throughput to 0 MB/sec while maintaining 99.9\% of benign traffic throughput.
\end{lightgraybox}

\subsection{RQ2: Defense Scalability}

\paragraph{Real-world workload traces}
We evaluate \tool using three datasets.
In DARPA OpTC, an attacker pivoted across 14 hosts beyond its initial compromise. 
Yatesbury benchmark consists of 13 attack scenarios in an e-commerce web-based application.
The LANL dataset collects benign activities from a real-world network of 17K hosts.
As LANL lacks malicious traces, we simulate realistic attacks (similar to~\cref{sec:eval_effectiveness}) for a comprehensive evaluation.

We use Distributed Internet Traffic Generator (D-ITG)~\cite{ditg} to generate the three workloads from our servers through the physical switch.
We evaluate three switch programs: 1) a forwarding-only baseline (no defense), 
2) a tagging switch program that updates the user’s \object, 
and 3) an enforcement switch program that enforces security decisions based on the \object.
We measure \tool's impact on benign traffic by comparing the flow completion time (FCT) under each of the three scenarios.

\cref{fig:datasets} shows the cumulative distribution function (CDF) of FCT for benign traffic.
We observe negligible differences between the baseline and \tool, because tagging and enforcement switches operate entirely in the data plane without involving the control plane.
Also, \tool inspects only the initial packet of each connection, forwarding subsequent packets without modification.
These findings confirm that \tool enforces per-user access scopes without impacting the performance of realistic workloads.

\paragraph{Large-scale network topology}
To evaluate \tool in complex and realistic topologies, we deploy three topologies: the social media network~\cref{fig:topology_social_network}, the media streaming service~\cref{fig:topology_media_service}, and the Stanford backbone network~\cref{fig:topology_stanford}.
For each topology, we repeat the attack scenarios (described in~\cref{sec:eval}) under two configurations: with distributed firewalls (iptables) and with \tool.
In every configuration, we pick a compromised host where the attacker begins and a target host that the attacker tries to reach.
The firewall baseline restricts access to the target to a subset of hosts, while in the \tool setup, policies explicitly limit the target to admin users, denying access from all others.

We run an attacker script that carries out the exploits and then probes all possible routes to the target, recording every successful route leading to the target.
With distributed firewalls, only direct connections from the attacker are blocked, covering just 3-5\% of all possible routes.
The attacker can still compromise intermediate hosts and bypass defenses to reach the target.
In another run, we deploy \tool in the topologies.
In contrast, when \tool is deployed, the attacker’s \objects are tracked across hosts, confining them to their authorized access scope.
As a result, the attacker cannot reach the target regardless of the pivoting route, increasing the percentage of blocked routes from 3-5\% to 100\%.

\begin{lightgraybox}
\textbf{Takeaways:} \tool scales to real-world workloads, imposing negligible latency on 3 tested datasets, and effectively increases the attack coverage by 33.3x across 3 real-world topologies compared to distributed firewalls.
\end{lightgraybox}

\begin{figure}[t]
    \centering
    \includegraphics[width=.85\linewidth]{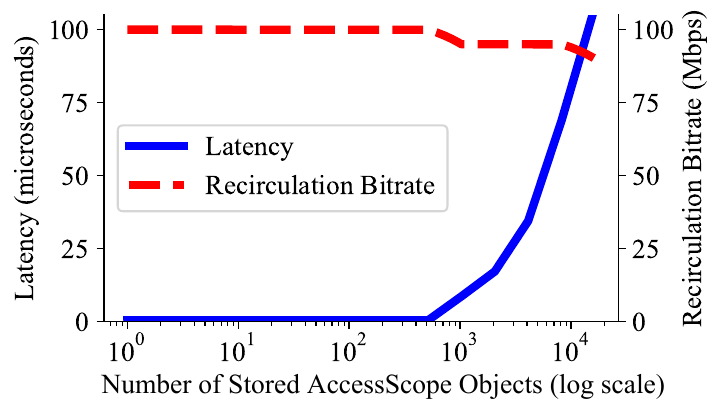}
    \caption{In-flight storage can support a sufficient number of \objects while incurring negligible latency.}
    \label{fig:eval_in-fight}
\end{figure}

\subsection{RQ3: \tool Overhead Impact}\label{sec:eval_system_impact}

\paragraph{In-flight \object overhead}
We evaluate \tool’s performance when handling multiple in-flight \objects by installing our tagging switch program on a physical Tofino switch.
We gradually inject \objects while initiating new connections that require them, continuing until the switch’s parser queue becomes saturated and packet drops occur.
To increase our in-flight storage, we apply port shaping to reduce the processing bitrate of \objects, allowing more \object packets to remain in the switch pipeline before re-entering the parser.

\cref{fig:eval_in-fight} shows the relationship between the number of stored \objects and the pipeline processing bitrate.
The bitrate declines gradually as more \objects are stored, dropping by $\sim$15\% when storing \objects for 16K users.
At this bitrate, tagging an incoming packet with its \object adds less than 110$\mu s$ of latency, which is negligible considering that the RTT in typical enterprise networks spans several milliseconds.
This \emph{only occurs} for the first packet of a new connection, as subsequent packets are forwarded without modifications.

\paragraph{Number of supported \objects}
Large-scale enterprise environments typically require support for up to 10K active users concurrently in a single cluster.
For instance, a single Google cluster~\cite{google_clusters1, google_clusters2} may include 1K-13K machines but usually serves 10K concurrent users.
As shown in~\cref{fig:eval_in-fight}, \tool efficiently stores up to 16K concurrent \objects, well above this requirement.
Thus, \tool provides sufficient capacity for large-scale deployments while accommodating future growth in the number of active users.

\paragraph{Number of supported active connections}
We leverage the feature of the P4 compiler to measure the capacity of \tool in handling active connections. 
We keep increasing the maximum number of active connections until the P4 compiler rejects our program and record the largest number supported. 
We find that up to 200K concurrent connections can be supported at the same time. 
This is more than the number of active connections found in Facebook frontend clusters, which range between 10K and 100K~\cite{silkroad2017}.

\begin{figure}[H]
    \centering
    \includegraphics[width=\linewidth]{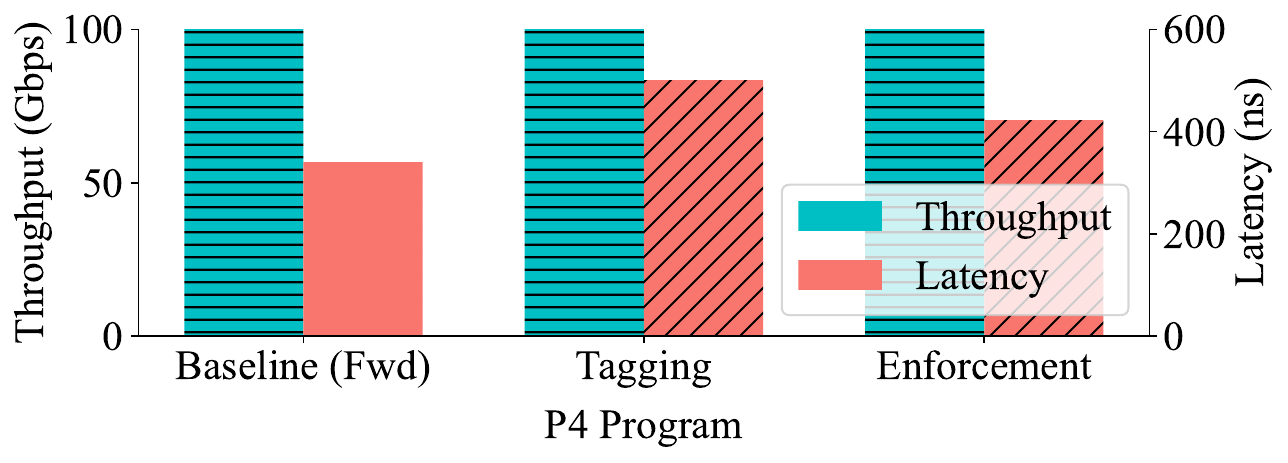}
    \caption{\tool tagging and enforcement primitives network performance compared to the forwarding baseline}
    \label{fig:latency_throughput}
\end{figure}

\paragraph{Throughput and latency}
We evaluate \tool’s impact on network performance by comparing it against a P4 program that performs simple packet forwarding.
We run the switch once with the tagging program and once with the enforcement program.
The comparison results are shown in~\cref{fig:latency_throughput}.
After successful compilation, the pipelined nature of the switch hardware ensures that the \tool’s P4 program operates at a rate of 99.9~Gbps per port, matching the forwarding baseline.
Also, \tool adds 110-160~ns of latency, which is negligible considering that the RTT in typical enterprise networks spans several milliseconds.

\begin{table}[t]
    \centering
    \small
    \begin{tabular}{lcc}
        \toprule
        \textbf{Kernel Function} & \textbf{Time ($ \mu $s)} & \textbf{\tool $\Delta$ ($ \mu $s)}  \\
        \midrule
         \verb|execve| & 432  & 5.1 \\
         \verb|clone| & 261 & 6.3 \\
         \verb|exit| & 11 & 0.7 \\
         \verb|TC egress| & 51 & 0.9 \\
         \verb|XDP ingress| & 32 & 0.3 \\
         \bottomrule
    \end{tabular}
    \caption{Original average time taken for kernel functions with their respective overhead introduced by our eBPF programs}
    \label{tab:ebpf_overhead}
\end{table}

\paragraph{eBPF programs overhead}
We evaluate the impact of our eBPF programs on a Linux server (kernel 5.15.0) by measuring both latency and memory overhead.
To quantify the runtime cost of propagating \object across system entities, we connected to a server using netcat, invoked 10K child processes, and measured the additional delay introduced by the eBPF programs.
To measure the cost of our TC and XDP programs, we generated 10K flows carrying \object and recorded the program's execution time.
Our results show that the programs add only 2-7$\mu s$ of latency per system call.
This overhead is negligible compared to the baseline runtime of {\cmtt execve()} and {\cmtt clone()} system calls, which we measured at 432$\mu s$ and 261$\mu s$, respectively.
We also analyze the memory requirements of storing \object in BPF maps.
With our 64~GB RAM system, Linux caps our system to about 1M processes and threads that can be concurrently created~\cite{process_thread_cap}.
With an \object with a 1028-bit width permission field, our BPF maps consume roughly 130MB of memory, which is insignificant given modern server resources.

\begin{lightgraybox}
\textbf{Takeaways:} \tool supports up to 16K users and 200K concurrent connections while maintaining line-rate performance and adding only 110-160~ns of latency. Its eBPF programs incur a negligible 2-7$~\mu s$ per system call.
\end{lightgraybox}

\subsection{RQ4: Comparison with Centralized User-Based Access Control Solutions}\label{sec:eval_comparison}

Recent SDN-based efforts~\cite{ethane, netview, fml} have introduced role- and attribute-based policy definitions.
Each new connection pauses until the controller verifies both endpoints and installs the required forwarding rules.
In our testbed, we set up an ONOS SDN controller, and we deploy both NetView~\cite{netview} and Ethane~\cite{ethane}. 
Since Ethane's code is not publicly available, we implement it as an SDN program.

\begin{figure}[t]
    \centering
    \includegraphics[width=.85\linewidth]{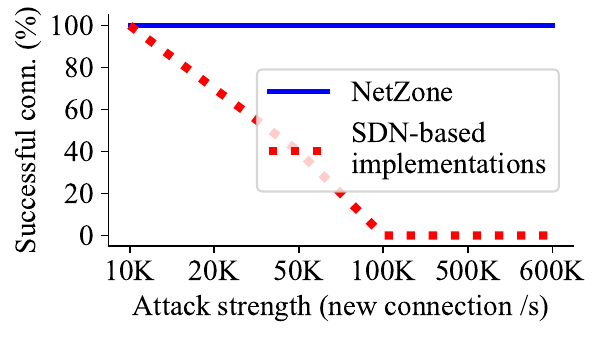}
    \caption{\tool in-network approach is resilient to saturation attacks compared to centralized systems}
    \label{fig:eval_server_based}
\end{figure}

\paragraph{Installing defense rules}
We evaluate how quickly the \tool switch-based design enforces access decisions for new connections.
In Ethane and NetViews, every new connection requires controller involvement.
The controller must process policy rules, verify the source and destination, and then install forwarding entries in the switches.
This controller-mediated process takes 10~ms-5~s, dominated by policy engine execution and delays in installing switch rules.
Such overhead is far too high for enterprise networks, where connection setup is expected to complete within milliseconds.

In contrast, \tool processes \object entirely within the data plane, eliminating controller processing and round-trip overhead.
Our tagging switches incur less than 110~$\mu s$ of delay on the first packet, while subsequent packets incur only 130~ns of latency. 
For enforcement, \tool can effectively enforce \objects within 140~ns.
This represents a 27-68x improvement over SDN-based approaches, suitable for high-throughput, low-latency enterprise requirements.

\paragraph{Saturating the control plane}
A centralized control plane quickly becomes a bottleneck under high connection rates.
An attacker can launch a saturation attack by initiating a large number of short-lived connections, forcing the controller to process a large number of setup requests and install rules. 
This not only increases per-connection latency but also overwhelms the controller’s processing capacity, leading to delayed or dropped legitimate connections.
As shown in~\cref{fig:eval_server_based}, SDN-based solutions are completely saturated beyond 100K packet/s.
In contrast, \tool enforces \objects entirely within the data plane, sustaining line-rate performance and remaining inherently robust against such saturation attacks.

\begin{lightgraybox}
\textbf{Takeaways:} \tool eliminates controller dependency, achieving a 27-68× speedup over centrlized systems such as Ethane~\cite{ethane} and NetView~\cite{netview}, while being resilient against control-plane saturation attacks.
\end{lightgraybox}

\begin{table}[t]
\caption{Hardware resource utilization}
\centering
\resizebox{\linewidth}{!}{
\begin{tabular}{@{}lcccc@{}}
\toprule
\textbf{\small Program} & \textbf{\small SRAM} & \textbf{\small TCAM} & \textbf{\small VLIW} & \textbf{\small Hash Units} \\ 
\midrule
{\small \tool Enforcement} & 17.3\% & 0.1\% & 2.1\% & 6\% \\ 
{\small \tool Tagging} & 21.1\% & 0.1\% & 2.9\% & 2.9\% \\ 
\midrule
\textbf{\small tna\_simple\_switch.p4} & 6.3\% & 9\% & 9.6\% & 0\% \\ 
\quad + {\small \tool Tagging} & 27.4\% & 9.1\% & 12.5\% & 2.9\% \\ 
\quad + {\small \tool Enforcement} & 23.6\% & 9.1\% & 11.7\% & 6\% \\ 
\midrule
\textbf{\small switch.p4} & 33.6\% & 31.6\% & 13.5\% & 19.4\% \\ 
\quad + {\small \tool Tagging} & 54.7\% & 31.7\% & 16.4\% & 22.3\% \\ 
\quad + {\small \tool Enforcement} & 50.9\% & 31.7\% & 15.6\% & 25.4\% \\ 
\bottomrule
\end{tabular}%
}
\label{tab:resource_usage}
\vspace{-1ex}
\end{table}

\subsection{RQ5: Switch Resources and Utilization}

\cref{tab:resource_usage} shows the switch resource utilization of both \tool tagging and enforcement programs as reported via the Intel P4 Insight tool.
Both \tool's programs consume a small footprint on shared resources (SRAM 17.3\%, TCAM 0.1\%).
As the tagging switch maintains \objects in-flight, it can store up to 16K \objects without incurring a footprint on the scarce switch memory.
Also, both programs make minimal use of VLIW units (for header modifications) and hash units (for in-network hash computations), leaving sufficient resources for additional P4 programs to run alongside \tool on the same switch.

In fact, we deployed \tool programs alongside two complex P4 programs: tna\_simple\_switch.p4 and switch.p4.
\cref{tab:resource_usage} demonstrates each of the P4 program and the total resources used after adding \tool.
These results show that \tool can be deployed alongside feature-rich programs on the same hardware switch while efficiently handling \objects away from scarce resources.

\begin{lightgraybox}
\textbf{Takeaways:} \tool uses only 17-21\% SRAM and 0.1\% TCAM, maintains minimal VLIW and hash unit utilization, and retains enough resources to coexist with other complex P4 programs on the same switch.
\end{lightgraybox}

\section{Discussion}
\label{sec:discussion}

\paragraph{eBPF security}
While eBPF provides a secure sandbox environment for running user-defined programs in the kernel, there are still some potential vulnerabilities with the current technology.
Recent zero-day vulnerabilities~\cite{ebpf-vul1, ebpf-vul2} enable attackers to execute arbitrary memory reads and writes, potentially compromising the integrity of our BPF maps.
Existing works have been proposed to enhance the safety verification of eBPF programs~\cite{ebpf-verify1, ebpf-verify2},
significantly reducing the eBPF vulnerabilities that may be used to exploit the kernel.
Such efforts further bolster the security of our eBPF programs.

\paragraph{Stolen user credentials}
\tool is not designed to detect attackers who steal user credentials (e.g., username and password) and authenticate to the IAM system to gain the victim’s access scope.
\tool can benefit from recent works~\cite{hopper, log2vec} that analyze login activities across the network to identify and flag suspicious authentication attempts.

\section{Related Work}
\label{sec:related_work}

\paragraph{In-network programmability}
Programmable switches are widely used in enterprise networks to offload various networking tasks~\cite{hula2016, redact2021, sonchack2018, redplane}.
A parallel line of work has explored in-network defenses against different attack types, such as covert channels~\cite{netwarden}, distributed denial-of-service~\cite{jaqen2021, smartcookie, poseidon2020}, data leakage~\cite{p4control}, link flooding~\cite{Jiarong2021, mew2023}, privacy threats~\cite{raven}, and RDMA vulnerabilities~\cite{bedrock}.
However, none of the existing in-network defenses enforce per-user, process-level access scopes at line rate.

\paragraph{Network isolation}
A large body of work focuses on isolating network traffic to contain breaches and limit lateral movement.
Traditional network virtualization techniques such as VLANs~\cite{vlan} and network slices~\cite{network_slice} provide coarse-grained separation at the subnet level.
More recent efforts~\cite{svlan, mondrian, psi, splended_isolation} aim to simplify virtualization definitions in large-scale environments.
Leveraging SDN, these systems introduce policy abstractions that allow administrators to specify VLAN requirements.
Another line of work~\cite{ethane, fml, netview} maps high-level role- and attribute-based policies into IP-based rules that mediate traffic.
While these mechanisms restrict communication between isolated networks, they still lack process-level visibility and cannot track users' movement, enabling attackers to find permitted paths among network subnets to laterally move.

\paragraph{eBPF for cybersecurity}
Existing efforts have leveraged eBPF to develop offensive tools and eBPF-based malware~\cite{ebpf-mal1, ebpf-mal2}. 
Other works have explored new attacks that leverage the nature of eBPF programs to target cloud containers~\cite{ebpfcross2023}. 
On the defensive side, several works~\cite{eaudit, ebpfaudit} have utilized eBPF to enhance OS logging.
\tool leverages eBPF to realize process-level capabilities that mediate access to remote resources without modifying the OS or the underlying kernel.

\section{Conclusion}
\label{sec:conclusion}

We presented \tool, an in-network, fine-grained defense that enforces per-user, access scopes to secure network access against lateral movement in enterprise networks.
\tool introduces \object, a lightweight, process-bound credential that encodes user permissions, propagates across hosts, and is enforced at line rate using programmable data planes.
Our design eliminates ambient host-level trust by binding permissions to individual processes and persisting them as users move across the network.

\bibliographystyle{plain}
\balance
\bibliography{ref}

\end{document}